# A Study on Recent Approaches in Handling DDoS Attacks

**Debajyoti Mukhopadhyay**[1, 2]

[1]Web Intelligence & Distributed Computing Research Lab
Green Tower, C-9/1, Golf Green, Calcutta 700095, India
debajyoti.mukhopadhyay@gmail.com

**Byung-Jun Oh**[2]    **Sang-Heon Shim**[2]    **Young-Chon Kim**[2]

[2]Advanced Communications & Networks Lab, Division of Electronics & Information Engineering
Chonbuk National University
561-756 Jeonju, Republic of Korea

*ABSTRACT*
*In this paper, we present a study on the recent approaches in handling Distributed Denial of Service (DDoS) attacks. DDoS attack is a fairly new type of attack to cripple the availability of Internet services and resources. A DDos attack can originate from anywhere in the network and typically overwhelms the victim server by sending a huge number of packets. Several remedial measures have been proposed by various researchers. This paper attempts to discuss the recent offerings to handle the DDoS attacks.*

## 1. INTRODUCTION
Internet security has been a concern for all the users. However, a very strange kind of incident took place in 1999 [24]. Another attack took place against Yahoo! in February 2000 [25]. Again on 20th October 2002, another DDoS attack took place, where 13 root servers responsible for providing Domain Name System service were affected. It caused seven of the thirteen root servers crippled [23].

A Distributed Denial of Service (DDoS) attack is an attack to prevent the users from using the resources of a victim's computer. It is a large scale attack in a co-ordinated fashion, which is typically launched indirectly with the help of other computers in the Internet. There are several kinds of DDoS attacks. There are two main classes of such attacks: (1) bandwidth depletion and (2) resource depletion attacks. In case of bandwidth depletion attack, the victim network is flooded with unwanted traffic that prevents legitimate traffic from reaching the victim computer. In the other case of resource depletion attacks, the attack is targeted to tie up the resources of the victim computer [27] [28].

A new kind of DDoS attack is known as DDoS *Reflector* attack. It is a kind of attack which is difficult to defend as the victim computer is flooded with traffic from other Internet servers, which were not even compromised. This attack exploits the SYN ACKs in response to the TCP SYN requests and other TCP packets. Basically, attackers misuse this acknowledging packet as a reflector.

The rest of the paper is organized as follows: in Section 2 we discuss further about the DDoS Reflector attack. In the following four sections, the preventive measures are discussed. In Section 3, the Rate Limit solution to handle DDoS attack is discussed; Defense by Offense is explained in Section 4; Active Filtering solution approach is discussed in Section 5; Section 6 talks about the IP Trace back approach; and the concluding remarks are in Section 7. There is also a Reference section at the end.

## 2. DDoS Reflector ATTACK

There are various kinds of DDoS attacks, though we will discuss a typical reflector attack in this section [29]. Any server that supports a protocol which replies with a packet after it has received a request packet can be misused as a reflector without the need for a server compromise [27].

In Figure 1, a reflector attack is shown. Here, the agents send their packets with the false source address, also known as spoofed address, to the victim's address, to innocent servers, which act as reflectors. The interesting thing is, the source addresses of the attack packets received by the victim are not spoofed, rather they are of the uncompromised servers. The set up of these attacks organize the "masters" and "agents" in such a way that, the rate of the packets and the size of the packets get amplified. As a result, it becomes very difficult to trace back an attack to the initiator of such an attack.

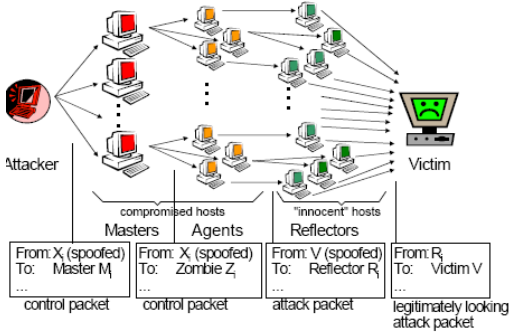

**Figure 1.** A DDoS Reflector Attack Scenario

## 3. RATE LIMIT

There are several approaches proposed by researchers to defeat DDoS attacks with the help of rate limit framework [12-15]. However, in this paper, we will discuss two such approaches.

We begin our discussion with the approach offered in the paper by Jing, et. al. [12]. It proposes the usage of three processes: a) attack detection, b) deciding the rate limit, and c) applying the rate limit to the attack traffic closer to the sources.

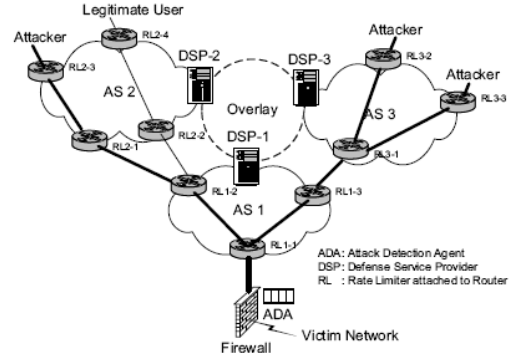

**Figure 2.** Architecture of $O^2$-DN

Figure 2 shows the three major processes of the $O^2$–DN architecture. These are: (1) Attack Detection Agent (ADA), (2) Defense Service Provider (DSP), and (3) Rate Limiter (RL).

ADA is installed as a software or hardware in the victim or the firewall. It is responsible for sending an alert and a defense request to the DSP, as soon as an attack is detected. The DSP is responsible for processing the defense service orders and provide defense services. When it receives a defense request, it verifies its authenticity to make sure that, it is not a new DoS attack. It then performs rate limit decision-making and sends rate limit commands to RL. The RL is responsible for limiting the rate of one specific flow. It also reports the approximate real-time rate information to the local DSP. RL is deployed by the Internet Service Provider and managed by the local DSP server in the same domain [12].

In another approach, Mahajan et. al. proposes a mechanism for detecting and controlling high bandwidth aggregates, which is typically resulted when a DDoS attack takes place [15].

An aggregate is a collection of packets from one or more flows with some similar properties. These properties could be: destination or source address prefix, certain application type like streaming video, TCP SYN packets, ICMP ECHO packets, etc. It can be very broad (e.g., TCP traffic) or very narrow (HTTP traffic) to a specific host.

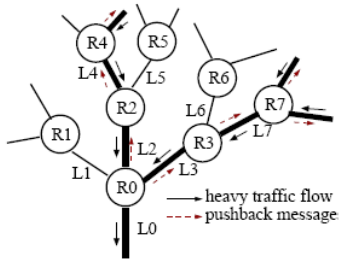

**Figure 3.** Pushback Approach

It proposes a pushback mechanism as shown in Figure 3. Pushback is a cooperative mechanism to control an aggregate upstream of network traffic. Here, the congested router sends commands to its adjacent upstream routers to rate-limit the aggregate. These are sent only to the contributing neighbors as those are responsible for sending major fraction of the aggregate traffic. Apart from saving upstream bandwidth through early dropping of packets that would have been dropped downstream at the congested router, this pushback method helps to focus rate-limiting on the attack traffic within the aggregate [15].

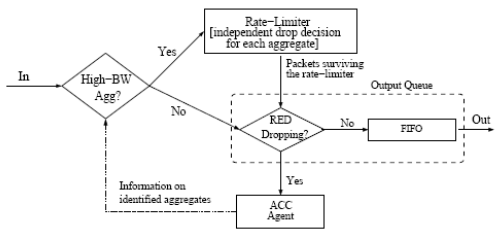

**Figure 4.** Architecture of an ACC-enabled Router

The proposed aggregate-based congestion control (ACC) architecture, which operates at the granularity of the aggregates is shown in Figure4. The packets of high-bandwidth aggregates pass through the rate-limiter. All packets dropped by RED are passed to the ACC Agent for identifying aggregates.

## 4. DEFENSE BY OFFENSE
There is an interesting proposal published by Walfish et. al. in which the remedial measure of DDoS attacks at application level is offered by sending higher volume of traffic[26].

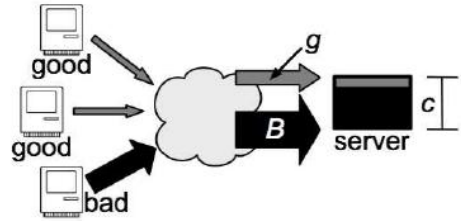

**Figure 5.** An Attacked Server Without Speak-up

The logic behind this scheme is not to just slow down the bad clients, but encouraging the good clients to send higher volume of traffic. It assumes that, if the bad clients are already using most of their upload bandwidth, by encouraging the clients to send higher volume of traffic will change in the volume of good clients only.

In Figure 5, diagram of an attacked server is shown without speak-up strategy. Here the bad clients have occupied a higher portion of the bandwidth. With a speak-up scheme in place, the bandwidth is now occupied more by the good clients as is shown in Figure 6.

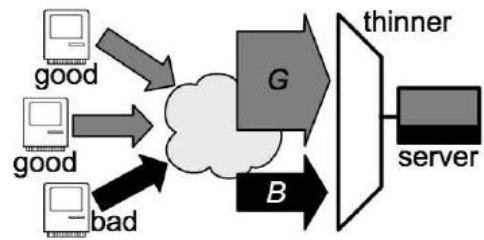

**Figure 6.** An Attacked Server With Speak-up

The speak-up needs a mechanism to measure the bandwidth. Its primary responsibility is to encourage the clients causing the clients to send more traffic. To implement these mechanisms, a front-end called thinner is used by the speak-ups. The thinner implements the encouragement and controls by sending requests to the server.

## 5. ACTIVE FILTERING
There are significant work done to protect victims from DDoS attacks by active filtering method [16-22].

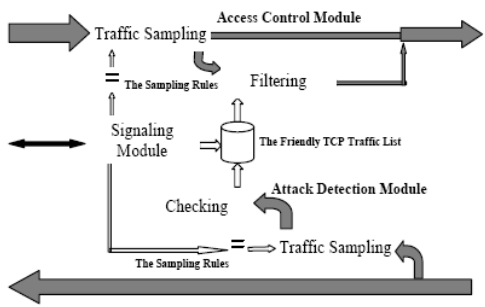

**Figure 7.** Architecture of the Gateway-based Defense System

We begin our discussion in this section with a Gateway Architecture by Xuan et. al. [16]. It proposes a defense system to detect attacks and control the traffic. It protects TCP friendly traffic occupying the main body of the Internet traffic. The gateways are deployed at various locations in the network to detect the attacks and perform the traffic access control as well. This architecture is shown in Figure 7.

In another work, Yaar et. al. offered a new idea, where they proposed a packet marking scheme based on Pi, and also a new filtering mechanisms [20]. The Pi DDoS defense scheme comprises of a packet marking algorithm, which encodes a complete Pi in each packet. It also comprises of a packet filtering algorithm, which determines how a DDoS victim uses this scheme effectively.

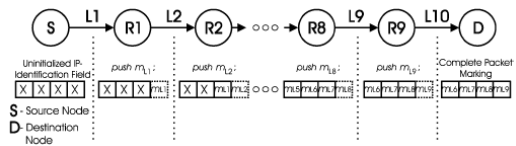

**Figure 8.** The Basic Stack Marking Scheme

The basic stack marking scheme is shown in Figure 8. It shows how the Pi mark evolves as the packet traverses R1 through R9.

Figure 9 shows an example of stack-based scheme with write ahead. This scheme allows the inclusion of markings from router R3. Stack marking eliminates the interaction between Pi- enabled routers and legacy routers that is present in TTL marking.

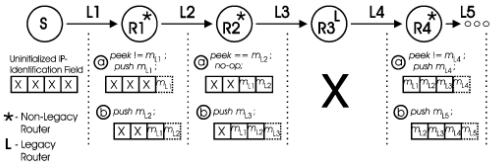

**Figure 9.** The Stack Marking Scheme with Write-ahead

It is assumed that each router knows IP address of the last-hop routers or hosts from which it receives packets. If it is also assumed that each router knows the IP address of the next-hop routers or hosts to which it is forwarding packets, then the router is capable of marking the packets on their behalf. All the router needs to substitute its own IP address for the last-hop IP address and the next-hop IP address for its IP address when calculating the bits to mark. This later marking is called write-ahead marking.

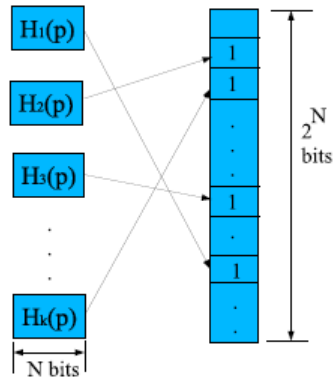

**Figure 10.** History-based IP Filtering Technique

There is another approach based on IP source address filtering proposed by Peng et. al. [22]. This approach utilizes history-based IP filtering for the edge router. Edge router is the router which provides access to the Internet to the domain of a sub-network, which this scheme is meant to defend. It uses a pre-built IP address database based on the history of all the legitimate

IP addresses, and accordingly admits the incoming packets.

Figure 10 shows the Bloom filter used for computing the *k* distinct IP address digests for each IP address using independent uniform hash functions. It uses the *N*-bit results to index into a $2^N$–sized bit array. The array is initialized to zeros and bits are set to one as packets are received. The validity is checked by computing the *k* digests on the IP address of an incoming IP packet and also checking the indicated bit positions. This test indicates that, if any one of them is zero, the IP address was not stored in the table.

## 6. IP TRACEBACK

Like the previous methods discussed in the earlier sections, there are also significant work done to protect victims from DDoS attacks by IP traceback method [1-11]. We will focus primarily on one such recent work.

Strayer et. al. proposes an architecture for multi-stage traceback to handle DDoS attacks[1]. This architecture is called *Stealthy Tracing Attackers Research Light TracE* (STARLITE), which is an extended version of *Source Path Isolation Engine* (SPIE) [30]. This STARLITE architecture constructs a prototype to integrate single packet traceback with stepping stone detection.

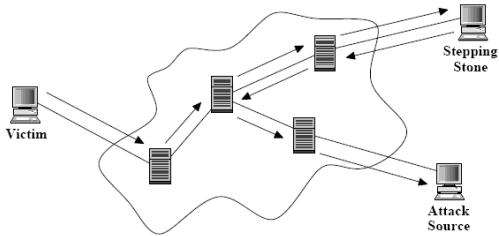

**Figure 11.** Trace across a Stepping Stone

In order to trace an attack path through laundering hosts, which are also called the *stepping stones* in this context, the chain of connections is discovered. In Figure 11, the chain of connections between an attacker and the victim is shown, which is called an extended connection.

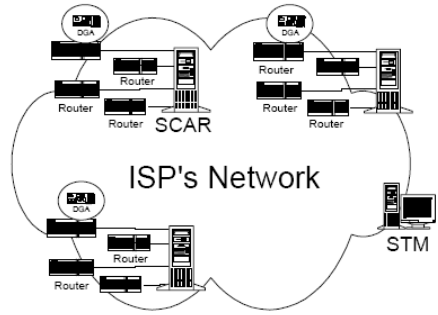

**Figure 12.** The SPIE Architecture

In Figure 12, the basic architecture of the SPIE system is shown. SPIE is a log-based traceback system. The basics of its tracebacking capabilities of the IP packets depend on auditing the network routers. The SPIE Traceback Manager (STM) is responsible for controlling the entire system.

The STARLITE system integrates the stepping stone detection concept into the SPIE system to utilize the rich communication infrastructure of SPIE.

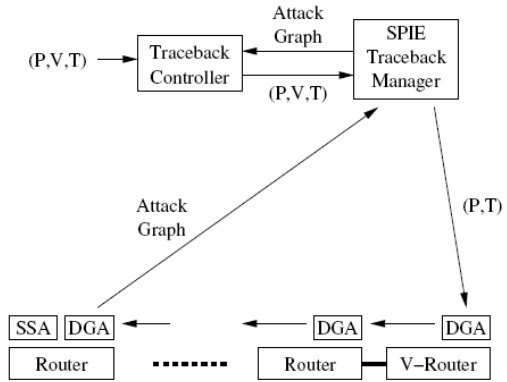

**Figure 13.** Multi-stage Trace – Step 1

The multi-stage trace proposed in STARLITE is shown in Figure 13. Here, *P* stands for a packet, *V* for a victim, and *T* for the time of attack. *DGA* is the Data Generation Agents in this diagram. *SSA* stands for the Stepping Stones Aggregator.

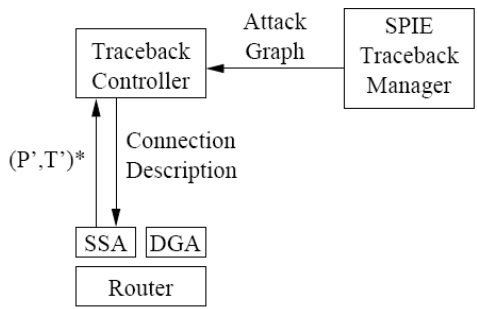

**Figure 14.** Multi-stage Trace – Step 2

The querying for stepping stones is shown in Figure 14. It shows how the Traceback Controller (TBC) receives the attack graph, identifies the router, and finally extracts the connection description from the traced packet.

Figure 15 shows the next step in the multi-stage traceback process. The Traceback Controller returns to the SPIE Traceback Manager to ask SPIE for constructing a new query based on the packet from the incoming connection and then continue the trace.

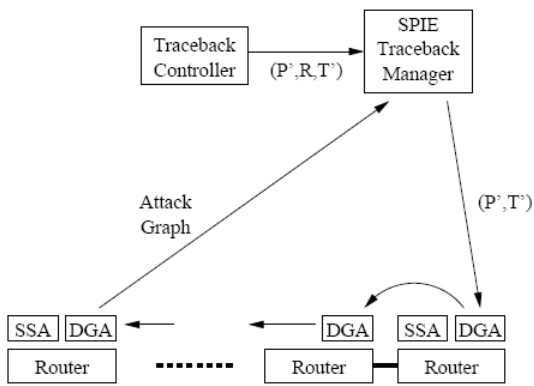

**Figure 15.** Multi-stage Trace – Step 3

## 7. CONCLUSIONS
Distributed Denial of Services attack poses great challenges to the researchers in the field of network security. It has already taken a heavy toll on many Internet based service providers in the world. There have been significant amount of work to tackle such DDoS attack with different kinds of approaches. In this paper, we have presented four major approaches that are being considered by the experts in this field. Perhaps it will be a hard and impossible task to discuss each and every published work in this field and propose the best solution. That's why we have kept the scope of the paper limited to just categorizing the existing solutions.

## 8. ACKNOWLEDGEMENTS
Authors of this paper gratefully acknowledge to the authors of the respective papers from where the diagrams have been borrowed. This work was supported in part by Chonbuk National University, by Ministry of Information & Communication of Republic of Korea under the IITA Visiting Professorship on IT Program, and by Techno India (West Bengal University of Technology), Kolkata, India. This work was done while the first author was at Chonbuk National University under the IITA Visiting Professorship on IT Program.